\documentclass[10pt]{article}
\usepackage[T1]{fontenc}
\usepackage[sc]{titlesec}
\usepackage{graphicx}
\usepackage{array}
\usepackage{multicol}
\usepackage{hyperref}
\usepackage[A4]{vmargin}

\newcommand{\dxdy}[2]{{\frac{\partial{#1}}{\partial{#2}}}}

\newcommand{\DxDy}

\def\rhobar{{\bar\rho}}

\def\Tbar{{\overline T}}

\def\gbar{{\bar g}}

\def\nubar{{\bar\nu}}

\def\etabar{{\bar\eta}}

\def\rhat{{\hat\mathbf{r}}}

\def\omegavec{\mathbf{\Omega}}
\def\grad{\mathbf{\nabla}}
\def\div{\mathbf{\nabla} \cdot}
\def\curl{\mathbf{\nabla} \times}
\def\vvec{\mathbf{v}}

\def\Bvec{\mathbf{B}}

\def\deg{^\circ}

\begin{document}
{\flushleft
  {\LARGE Constraints on the Magnetic Field Strength of HAT-P-7 b and other
    Hot Giant Exoplanets}
  \smallskip
  
  T.M. Rogers$^{1,2}$
  \smallskip
    
  \begin{trivlist}
  \item $^1$ Department of Mathematics \& Statistics, Newcastle University, UK
  \item $^2$ Planetary Science Institute, Tucson, AZ, 85721, USA
  \end{trivlist}
}

\section*{Summary}
Observations of infrared and optical light curves of hot Jupiters have
demonstrated that the peak brightness is generally offset eastward
from the substellar point [1,2].  This observation
is consistent with hydrodynamic numerical simulations that produce
fast, eastward directed winds which advect the hottest point in the
atmosphere eastward of the substellar point [3,4].
However, recent continuous Kepler measurements of HAT-P-7 b show that
its peak brightness offset varies significantly in time, with
excursions such that the brightest point is sometimes westward of the
substellar point [5].  These variations in brightness
offset require wind variability, with or without the presence of
clouds.  While such wind variability has not been seen in hydrodynamic
simulations of hot Jupiter atmospheres, it has been seen in
magnetohydrodynamic (MHD) simulations [6].  Here we show that
MHD simulations of HAT-P-7 b indeed display variable winds and
corresponding variability in the position of the hottest point in the
atmosphere. Assuming the observed variability in HAT-P-7 b is due to
magnetism we constrain its minimum magnetic field strength to be 6\,G.
Similar observations of wind variability on hot giant exoplanets, or
lack thereof, could help constrain their magnetic field
strengths. Since dynamo simulations of these planets do not exist and
theoretical scaling relations [7] may not apply, such
observational constraints could prove immensely useful.

\begin{multicols}{2}
\section*{Main Text}
To demonstrate magnetic effects on the winds of HAT-P-7 b, we simulate
the atmosphere of a hot giant exoplanet with parameters similar to
HAT-P-7 b using a spherical, three-dimensional (3D), anelastic MHD
code [8,6].  We start with a hydrodynamic simulation of
HD209458 b in terms of gravity, radius and rotation, but with a mean
temperature (2200K) and temperature differential (1000K) of HAT-P-7 b
(temperature and magnetic diffusivity profiles are shown in
Supplementary Figure~1).  The strong day-night temperature
differential drives strong eastward atmospheric winds, consistent with
previous simulations [9,10].  This simulation is run for
$\sim$100 rotation periods before a magnetic field is added after
which both the hydrodynamic and MHD simulations are run for an
additional ~280 rotation periods. Details of the numerical code and
simulation can be found in the Methods section.

  The extreme temperatures of HAT-P-7 b give rise to significant
  thermal ionization of alkali metals [11,12], which
  leads to coupling of the atmosphere to the deep seated magnetic
  field [13] and could also lead to an atmospheric
  dynamo [14].  The Lorentz force arising from this magnetic
  interaction disrupts the strong eastward directed atmospheric winds
  typically seen in hydrodynamic simulations, leading to variable and
  even oppositely directed winds [6].  Figure~1 shows a
  time-snapshot of magnetic field lines in the simulation looking onto
  the eastside terminator (a corresponding video of its complex
  evolution and variability is available in the Supplementary
  Material). The zonal-mean zonal-wind, averaged within 17$\deg$ of
  the equator and over the upper 1 mbar of the simulated domain, as a
  function of time is shown in Figure~2, along with the position of
  the hottest point in the atmosphere (also determined by an average
  over the same latitudes and height).  There we see that, the
  hydrodynamic model retains a strong, eastward jet and associated
  positive hotspot displacement throughout the simulation (dotted line
  in Figure 2a and 2b).  When a magnetic field is added, the zonal
  winds slow dramatically, reverse and then settle into an oscillatory
  pattern, with a timescale of $\sim$$10^6$\,s, consistent with the
  Alfven time ($\tau_A=\sqrt{4\pi\rho}\lambda/B$) of the imposed 10\,G
  field and is of the same order as the timescale of variability
  observed in HAT-P-7 b [5].  Variability in the hot
  spot displacement, including negative offsets, is seen on a similar
  timescale.
  
\end{multicols}\begin{figure}[h!]\centering
  \includegraphics[width=0.9\textwidth]{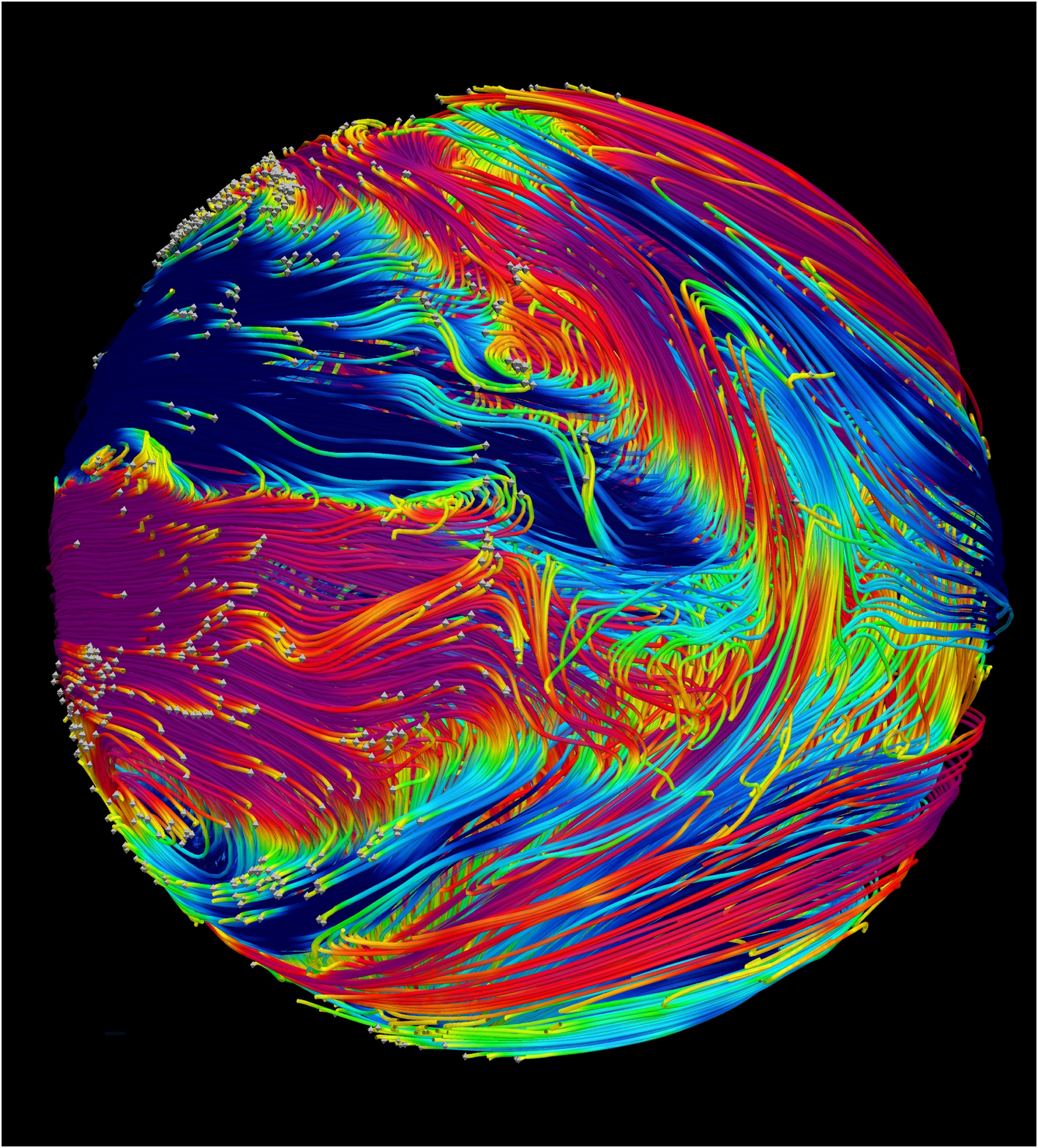}
  \caption{Magnetic Field lines in the atmosphere of a Hot Giant
    Exoplanet.  Time snapshot of magnetic field lines in the numerical
    simulation of a hot Jupiter atmosphere (a model of HD209458 b but
    with a temperature structure similar to HAT-P-7 b). Magnetic field
    lines are color-coded to represent the azimuthal (toroidal) magnetic
    field with blue representing negative directed field (saturated at
    -50\,G) and magenta positive (saturated at 50\,G), with green and yellow
    ranging from -5\,G to 5\,G, respectively.  The vantage point is looking
    onto the east-side terminator.}
\end{figure}\begin{multicols}{2}
\end{multicols}\begin{figure}[h!]\centering
  \includegraphics[width=0.9\textwidth]{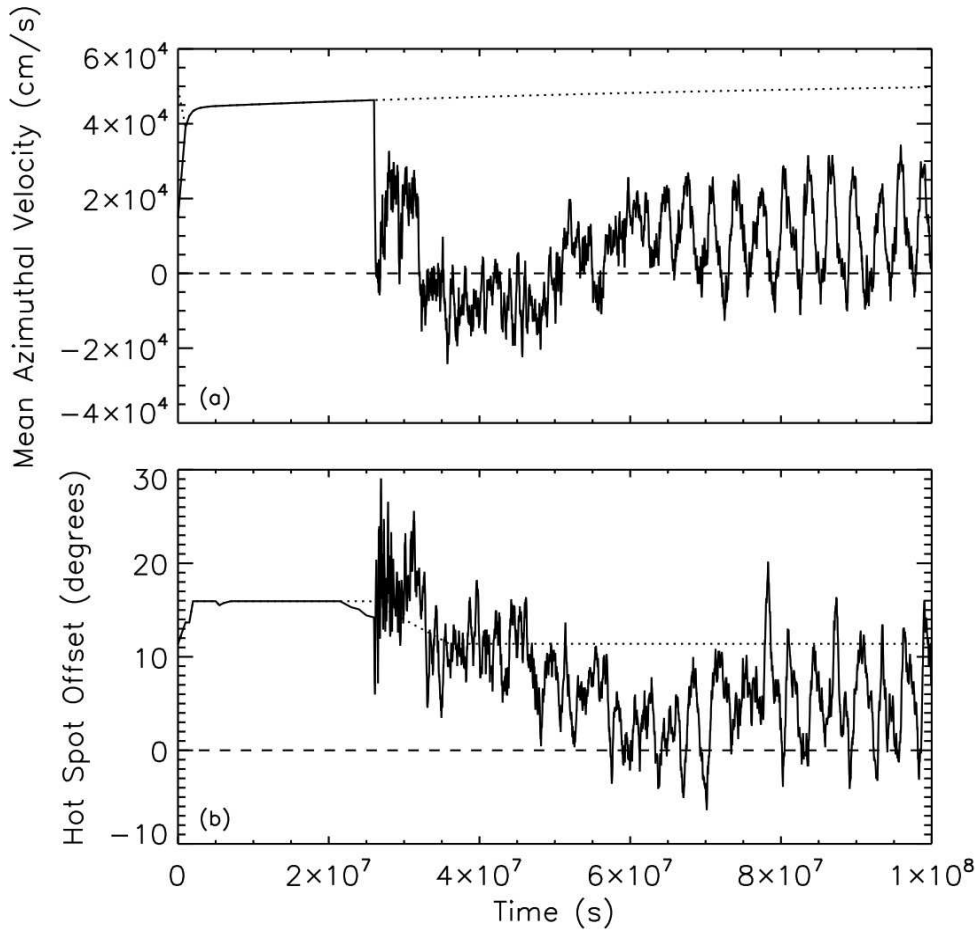}
  \caption{Atmospheric Dynamics of Simulated Hot Giant Exoplanet.] (a)
  Zonal Mean Zonal Wind in the hot Jupiter atmosphere averaged over
  17$^{\circ}$ around the equator and over the upper 1 mbar of the
  simulated domain.  Dotted line shows the winds in the hydrodynamic
  model while the solid line shows the winds in the MHD model.  (b)
  Displacement of the hottest point of the atmosphere from the
  substellar point, at the same location and averaged as in (a).
  Similar to the mean winds, the hot-spot displacement in the MHD
  model (solid line) shows strong variability with excursions to
  points west of the substellar point. Hydrodynamic models show a
  stable, positive offset.}
\end{figure}\begin{multicols}{2}
Both the hydrodynamic and MHD models have more positive hot-spot
displacements than the observations.  This is expected given that the
waves that force super-rotation can propagate further in HD209458 b
than HAT-P-7 b before being damped [15].  Therefore, we expect
a hydrodynamic model with HAT-P-7 b's gravity and rotation rate would
show reduced hot-spot displacements compared to HD209458 b and we
indeed find this (see Figure~3).  While this magnetic model has some
uncertainties (enhanced viscosity, crude radiative transfer), it
naturally explains the bright spot excursions as due to changes in the
thermal structure of the planet caused by variable winds.  In this
model clouds may not be necessary as HAT-P-7 b is hot enough that even
the optical signal could be dominated by thermal emission. Moreover,
this model may also explain the timescale of the observed fluctuations
as due to Alfven waves.  At the very least, it can provide the wind
variability needed for models requiring clouds [5].

The effect of magnetism on zonal winds depends on the ratio of the
magnetic to inertial terms in the momentum equation, which can be
approximated as the ratio of magnetic to wave timescales
$\tau_{mag}/\tau_{wave}$, where $\tau_{mag}=4\pi\rho\eta/B^2$ and
$\tau_{wave}=L/\sqrt{gH}$.  Here $\rho$ is the density, $\eta$ is the
magnetic diffusivity, $B$ is the magnetic field strength, g is the
gravity, L is the characteristic length scale of the horizontal flow
and H is the depth of the atmosphere [11,15].  As magnetic
effects are increased, either through increased magnetic field
strength or increased conductivity, their effect on atmospheric zonal
winds progressed from little to no effect (when $\tau_{mag} >
\tau_{wave}$), to oscillatory winds (when $\tau_{mag} \sim
\tau_{wave}$) to completely reversed (westward) winds (when
$\tau_{mag} < \tau_{wave}$) [6]. Assuming the variable winds
observed on HAT-P-7 b are due to magnetism and applying the
oscillatory wind condition we find $B\sim
\sqrt{4\pi\eta\rho/\tau_{wave}}$. Using the nightside value of $\eta$,
we find that HAT-P-7 b must have a minimum field strength of $\sim$6\,G.
This value is consistent with the theoretical scaling relation based
on the Elsasser number [16] ($\Lambda= 2\rho\Omega/\mu_0 \eta
\sim 1$) and with the upper limit placed on WASP-12
b [17], if we were to assume it had a similar field
strength.

To check this constraint, we ran additional models of HAT-P-7 b, with
the appropriate rotation, gravity, size and
temperature [18].  The temperature and magnetic diffusivity
profiles for this model can be seen in Supplementary Figure 2.  After
running a hydrodynamic model for ~140 rotation periods a magnetic
field was added and run an additional 15 rotation periods.  We show
the hot spot displacement for those models in Figure 3.  We see that
the hydrodynamic model (black line) has a steady hot-spot displacement
of 2.8$\deg$.  The MHD model with a 3\,G field (red line) shows a
similar, stable hot spot displacement.  However, both the 10\,G (blue
line) and 20\,G (orange line) model show wind variability ranging from
$\sim$$-15\deg$--$20\deg$.  This range of displacement is more
consistent with brightness variations observed (which range from
$\sim$$-25\deg$--$25\deg$). However, clouds could also play a role in
enhancing the large displacements observed by
Kepler [5].

In our simulations wind variability sets in between 3 and 10\,G,
consistent with the 6\,G lower limit based purely on a simple timescale
analysis.  If we had used the dayside magnetic diffusivity in the
estimate the lower limit would have been $\sim$0.6\,G, inconsistent with
our follow up models, which show no variability at 3\,G.  This estimate
depends only on the winds being variable and is independent of whether
or not clouds are needed to explain the exact range of variability
seen.  Although these models are consistent with the 6\,G lower limit,
on this timescale we see no completely reversed winds and therefore,
conclude that we can only place a \textit{ lower} limit on the field
strength.  While it may be possible to hone this constraint with more
simulations, its likely not worthwhile given the other limitations of
these simulations.

The continuous observations of HAT-P-7 b [5] were unique
in that previous optical and infrared observations have generally only
provided this measurement at a single epoch. The exception is the
multiple epoch Spitzer observations of HD189733 b [19]. That
work showed a fairly stable, positive offset.  This lack of
variability is consistent with little or no magnetic effects in
HD189733 b, a plausible conclusion given that the low temperatures of
HD189733 b would require unrealistic magnetic field strengths of
$\sim$100--1000\,G to cause variability.  In general, we would expect to
see wind variability in objects where field-flow coupling is strong
(as measured by the ratio of magnetic and wave timescales).
Therefore, we would predict variability may also be found in other hot
giant exoplanets, such as WASP-19b or WASP-12b.

While long timeline or multiple epoch observations of hot Jupiters'
phase curves have not been carried out for many objects, such a
campaign coupled with MHD models of those planets' atmospheres could
be used to place constraints on the magnetic field strengths of hot
Jupiters.  Such constraints are rare [20] and would be
useful for dynamo theory, planetary evolution and interpretations of
star-planet magnetic interactions [17].  As recently
shown [5], these types of constraints are already
possible with Kepler but will become more readily available with
upcoming space missions such as JWST, CHEOPS, TESS and PLATO.  In
particular, JWST will be able to measure infrared phase curves
directly, thus testing this theory without the complication clouds
might add to optical curves.
\end{multicols}

\begin{figure}\centering
  \includegraphics[width=0.9\textwidth,trim=13mm 52mm 10mm 67mm,clip=true]{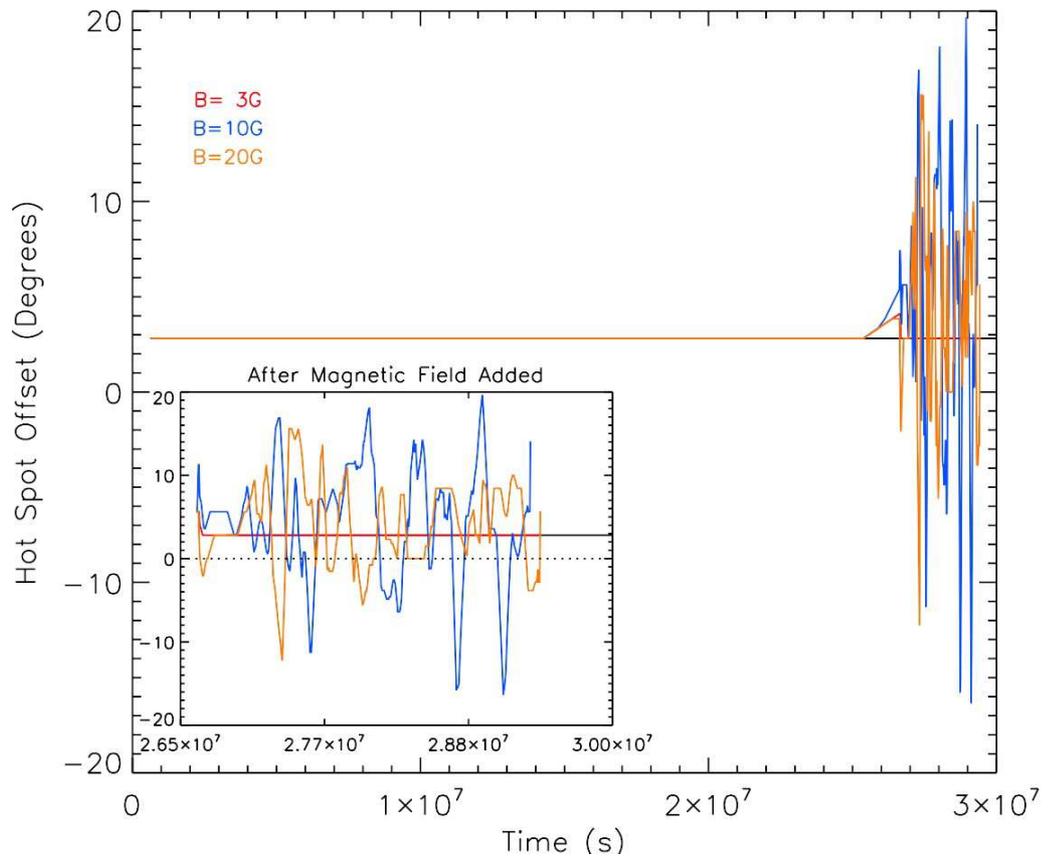}
  \caption{Hot spot displacement of Simulated HAT-P-7b.] Hottest point
  in the atmosphere, calculated after a latitudinal average around
  17$\deg$ of the equator and the upper 1mBar of the simulated domain,
  as a function of time.  Black line shows the hydrodynamic model
  (barely visible under the other lines), red line is for a 3\,G field,
  blue line is for a 10G field and orange line is for a 20\,G field.
  The dotted line shows the sub-stellar point.  The inset shows the
  time behavior after the magnetic field is added.}
\end{figure}

\section*{Addendum}

\subsection*{Acknowledgments} T.~R. thanks J.~N.~McElwaine, and
G.~Glatzmaier for helpful discussions leading to this manuscript and
J.~Vriesema for help with the graphics. Funding for this work was
provided by NASA grant NNX13AG80G and computing was carried out on
Pleiades at NASA Ames.

\subsection*{Author Contributions} T.R. carried out all work related to this manuscript.

\subsection*{Requests and Correspondence} 
Correspondence and requests for materials should be addressed to
T.~M.~Rogers\\
\href{mailto:tamara.rogers@newcastle.ac.uk}{tamara.rogers@newcastle.ac.uk}

\section*{methods}
We solve the magnetohydrodynamic (MHD) equations in three-dimensional (3D),
spherical geometry in the anelastic approximation\cite{rk14}.  The model solves the following equations:   
\vspace{0.15cm}
\begin{eqnarray}
\div (\rhobar \vvec) &=& 0
\\
\div \Bvec &=& 0
\\
\rhobar {\partial \vvec \over \partial t} +\div(\rhobar \vvec \vvec) &=&
- \grad p
- \rho \gbar \rhat
\\ \nonumber
&& + 2 \rhobar \vvec \times \omegavec
+ \div \left[2 \rhobar \nubar
(E - {1 \over 3} ( \div \vvec ) \mathbf{1})\right]
+ {1 \over \mu_0} ( \curl \Bvec ) \times \Bvec
\\
\lefteqn{\dxdy{T}{t}+(\mathbf{v}\cdot\nabla){T}=-v_{r}\left[\dxdy{\overline{T}}{r}-(\gamma-1)\overline{T}h_{\rho}\right]+(\gamma-1)Th_{\rho}v_{r}+}
\\ \nonumber
&& \gamma\overline{\kappa}\left[\nabla^{2}T+(h_{\rho}+h_{\kappa})\dxdy{T}{r}\right] + \frac{T_{eq}-T}{\tau_{rad}}+\frac{\eta}{\mu_{o}\rho c_{p}}|\nabla \times \mathbf{B}|^{2}
\end{eqnarray}
Equation (1) represents the continuity equation in the anelastic
approximation [21,22].  This approximation allows some level
of compressibility by allowing variation of the reference state
density, $\rhobar$, which varies in this model by four orders of
magnitude.   Equation (2) represents the conservation of magnetic
flux.  Equation (3) represents conservation of momentum including
Coriolis and Lorentz forces. Equation (4) represents the energy
equation including  a forcing term to mimic stellar insolation (fourth term on right hand
side, where T$_{eq}$
is the equilibrium temperature) and Ohmic heating (fifth term on right hand side). The radiative timescale in the
Newtonian forcing term, $\tau_{rad}$ is a function that varies between
10$^4$ s at the outermost layers 10$^6$ s at the lowest
layers. All other variables take their usual meaning [6].
  
The magnetic diffusivity $\eta$ (inverse
conductivity) is a function of all space.  If we separate the magnetic diffusivity into
a mean ($\etabar$) and fluctuating ($\eta'$) component:  
\begin{equation}
\eta\left(r,\theta,\phi\right)=\etabar\left(r
\right)+\eta'\left(r,\theta,\phi \right)
\end{equation}
where r, $\theta$ and $\phi$ are the radius, colatitude and longitude,
respectively.  The magnetic induction equation becomes 
\begin{equation} 
{\partial \Bvec \over \partial t} =
\curl ( \vvec \times \Bvec-\eta'\curl\Bvec )
- \curl (\etabar \curl \Bvec)
\end{equation}
Equation (6) is solved along with Equations (1)-(4).  

The magnetic diffusivity (5) is calculated from the initial 
temperature profile given by: 
\begin{equation}
T_{eq}\left(r,\theta,\phi\right)=\Tbar(r)+\Delta T_{eq}(r)
\cos\theta\cos\phi
\end{equation}
where $\Tbar(r)$ is mean reference state temperature and  $\Delta T_{eq}$ is the  specified
day-night temperature difference, here set to 1000K and 
which is extrapolated logarithmically from the surface to 10 bar.  
Using this temperature profile, the magnetic diffusivity is calculated
as[23]: 
\begin{equation}
\eta\left(r,\theta,\phi\right)=230\frac{\sqrt{T}}{\chi_e}
\end{equation}
and $\chi_e$ is the ionization fraction.  The ionization fraction is
calculated at each point using a form of the Saha equation taking into account all
elements from hydrogen to nickel and typical elemental abundances[24].

The model presented in Figure~1 and~2 is the model for HD209458
b [25] with 800K added at each vertical level and with an imposed
day-night temperature variation of 1000K.  The rotation rate, radius
and gravity are all that of HD209458 b.  The temperature and
diffusivity profiles of this model are shown in Supplementary Figure
1. While this model is clearly not HAT-P-7 b it has a temperature (and
thus, conductivity) structure similar to that expected for HAT-P-7 b.
Since it is the temperature (conductivity) structure that dominates
the MHD behavior of the atmosphere, this model is probably a faithful,
albeit imperfect, representation of the atmospheric dynamics in
HAT-P-7 b.  The model has 10G poloidal field imposed at the bottom
boundary.

The model presented in Figure 3 is a model of HAT-P-7 b using an atmosphere
model for HAT-P-7 b [18].  The temperature and magnetic
diffusivity profiles can be seen in Supplementary Figure 2.  Here,
dipole fields of 3G, 10G and 20 G are imposed at the bottom boundary
to mimic the deep seated dynamo field.  

Both the models presented have more complex dynamics than those found
previously [26,6] because these include a magnetic diffusivity
(conductivity) that is a function of all space.  This led to more
complex field-flow interactions, particularly at the terminators
(both) and even led to an atmospheric dynamo [14].  Although it
was not included here a time-dependent conductivity could further
complicate matters, particularly with regard to the thermal structure
of the atmosphere.  Currently, we see more Ohmic heating on the night
side of the planet, which leads to reduction of the day-night
temperature gradient.  Naively, if we allowed this to react back on
the flow and conductivity we would expect decreased wind driving and
increased field-flow coupling.  That is, we might expect wind
variability at even lower magnetic field strengths.

\paragraph{Data Availability Statement} The data that support the plots
within this paper and other findings of this study are available
from the corresponding author upon reasonable request.

\bibliographystyle{naturemag}




\end{document}